\documentclass[preprint,12pt]{elsarticle}
\usepackage{algorithm}
\usepackage{tikz}
\usetikzlibrary{fit,calc,positioning}
\usepackage{algpseudocode} 
\usepackage[braket]{qcircuit}
\usepackage{caption}
\usepackage{amsmath,amssymb,amsfonts}
\usepackage{braket}
\usepackage{graphicx}
\usepackage{booktabs}
\usepackage{threeparttable}
\usepackage{subcaption}
\usepackage{textcomp}
\usepackage{makecell}
\usepackage{xcolor}
\usepackage{hyperref}
\usepackage{multirow}
\newcommand{\modelname}{DALNet}

\def\BibTeX{{\rm B\kern-.05em{\sc i\kern-.025em b}\kern-.08em
    T\kern-.1667em\lower.7ex\hbox{E}\kern-.125emX}}

\usepackage{amsmath,algorithm,tabularx}
\makeatletter
\newcommand{\multiline}[1]{%
  \begin{tabularx}{\dimexpr\linewidth-\ALG@thistlm}[t]{@{}X@{}} 
    #1
  \end{tabularx}
}
\makeatother
\usepackage[font=small,labelfont=small]{caption}
\usepackage{url}    
\usepackage{amssymb}
\usepackage{nomencl}

\usepackage[margin=1in]{geometry}

\journal{IJEPES}
\makeatletter
\def\ps@pprintTitle{%
    \let\@oddhead\@empty
    \let\@evenhead\@empty
    \let\@oddfoot\@empty
    \let\@evenfoot\@empty
}
\makeatother
\begin{document}

\begin{frontmatter}

\title{DALNet: A Denoising Diffusion Probabilistic Model for High-Fidelity Day-Ahead Load Forecasting } 

\author[label1]{Han Guo}
\author[label1]{Ding Lin\corref{cor1}}\cortext[cor1]{D. Lin is the corresponding author (email:dinglin@smu.edu)}
\affiliation[label1]{organization={Electrical and Computer Engineering Department, Southern Methodist University},
            city={Dallas},
            state={TX 75205},
            country={USA}}

\begin{abstract}

Accurate probabilistic load forecasting is crucial for maintaining the safety and stability of power systems. However, the mainstream approach, multi-step prediction, is hindered by cumulative errors and forecasting lags, which limits its effectiveness in probabilistic day-ahead load forecasting (PDALF). To overcome these challenges, we introduce DALNet, a novel denoising diffusion model designed to generate load curves rather than relying on direct prediction. By shifting the focus to curve generation, DALNet captures the complex distribution of actual load time-series data under specific conditions with greater fidelity. To further enhance DALNet, we propose the temporal multi-scale attention block (TMSAB), a mechanism designed to integrate both positional and temporal information for improved forecasting precision.  Furthermore, we utilize kernel density estimation (KDE) to reconstruct the distribution of generated load curves and employ Kullback–Leibler (KL) divergence to compare them with the actual data distribution. Experimental results demonstrate that DALNet excels in load forecasting accuracy and offers a novel perspective for other predictive tasks within power systems.

\end{abstract}

\begin{highlights}
\item A novel denoising diffusion model, DALNet, is proposed for probabilistic day-ahead load forecasting.
\item  The model shifts the paradigm from direct prediction to curve generation, overcoming cumulative errors and forecasting lags.
\item  A new Temporal Multi-scale Attention Block is designed to capture both positional and temporal information in load data.
\item Kernel Density Estimation and KL divergence are used to rigorously validate the model's ability to learn the true data distribution.
\end{highlights}

\end{frontmatter}
\section{Introduction}
Day-ahead electrical load forecasting plays a key role in balancing electricity supply and demand, offering valuable insights for planning, construction, and operational decision-making in power systems. With the increasing integration of renewable energy, flexible resources, and advancements in electricity markets and demand response mechanisms, load forecasting has become more complex due to heightened volatility and uncertainty. In this evolving landscape, accurate and reliable probabilistic day-ahead load forecasting serves as a crucial tool for mitigating uncertainties and ensuring safe operational
control within the power system.

Currently, day-ahead forecasting can be broadly categorized into three methods. The first method is rolling forecasting, the second involves building individual models for each time point, and the third is multi-step prediction. 
The earliest methods for rolling forecasts employed statistical models like the autoregressive integrated moving average  model \cite{ARIMA} and the multiple linear regression  model \cite{MLR}. These linear models often need to catch up in capturing nonlinear relationships accurately. To address this issue, researchers have explored machine learning models such as support vector machines \cite{dai2020hybrid} and random forests \cite{prasad2019designing}. Despite their advantages, traditional machine learning algorithms sometimes need help modeling complex nonlinear relationships and managing large datasets effectively.
Consequently, neural networks have gained significant attention due to their superior performance in handling complex nonlinear mappings. For instance, Bi-directional long short-term memory (Bi-LSTM) \cite{yang2024attention} and feedback neural network \cite{IFNN} have been employed for rolling forecasts. 

Rolling forecasts often focus on point predictions and face the issue of error accumulation. Researchers have started modeling each time step individually for day-ahead load forecasting to address these challenges. \cite{MLR24} developed 24 multiple linear regression models, one for each hour of the day, for day-head forecasting. In addition, scholars often use neural networks to implement this method, such as the attention mechanism \cite{zhou2025photovoltaic}. Furthermore, the second day-ahead load forecasting method can be easily extended to probabilistic forecasting methods through techniques like quantile regression (QR) \cite{QRbook} and KDE \cite{KDE}. Y. Wang \textit{et al.} \cite{QLSTM} extended LSTM-based point forecasting to the quantile regression LSTM to handle the non-stationary and stochastic features of individual consumers. Taking the uncertainty of low-voltage load data into account to improve the accuracy of the probabilistic forecasting, Z. Cao \textit{et al.} \cite{CaoZhaoJing} proposed a hybrid ensemble learning model based on the deep belief network. X. Liu \textit{et al.} \cite{Zijun} proposed an ordinary differential equation network combined with QR to capture the uncertainties. M. Sun \textit{et al.} \cite{BayesianTheory} proposed a probabilistic day-ahead net load forecasting method that combined Bayesian theory and LSTM to capture the epistemic and aleatoric uncertainty of load data.

Although the second method can avoid the accumulation of prediction errors, independent modeling of each time point needs to pay attention to the relationships between different times of the day, which is a significant limitation for capturing intra-day load patterns with apparent regularities. The third method, multi-step prediction, has emerged to address this issue and has become the mainstream forecasting approach. By utilizing a transformer to capture the periodicity in load data, B. Jiang \textit{et al.} \cite{Tranformer} achieved multi-step forecasting. To improve the accuracy of day-ahead load forecasting of the Transformer, K. Qu \textit{et al.} \cite{forwardformer} proposed a novel model  excelling in predictions for special days such as weekends and holidays. For day-ahead probabilistic load forecasting, \cite{ensemble} first utilized ensemble learning for point forecasting, followed by Markov Chain Monte Carlo methods to generate the distribution of the forecasted day and achieve probabilistic forecasting. \cite{Fussion} leveraged a hybrid approach, combining Convolutional Neural Networks  and Gated Recurrent Units  for load forecasting, followed by QR to achieve probabilistic prediction. Gaussian Mixture Model  \cite{GMM} and KDE \cite{KDEproba} are also used to achieve probabilistic forecasting.

Although researchers widely apply multi-step prediction methods, these methods often suffer from a forecasting lag—that is, the predicted load curve tends to trail behind the actual trajectory, particularly during periods of rapid load variation (e.g., morning or evening ramps) \cite{LatencyProblem}. This effect is a well-known limitation of recurrent and autoregressive forecasting models, which smooth sharp transitions and introduce phase delays in their outputs. Therefore, new prediction methods are needed to address the shortcomings of error accumulation, lack of consideration for intra-day correlations, and the forecasting lag issue in the three methods above. To this end, we propose a prediction method based on conditional diffusion probabilistic models (cDDPM), a generative model, to generate load data for day-ahead forecasting directly. cDDPMs have demonstrated remarkable abilities in learning complex, high-dimensional data distributions and generating realistic samples. By adding Gaussian noise in a forward process and then learning to reverse it, cDDPMs can effectively recover the original data distribution, making them highly effective for generative tasks. The decision to use cDDPMs stems from their unique capabilities. Unlike rolling forecasting, cDDPMs can generate an entire curve without depending on the generated value from the previous time step, which avoids error accumulation. Additionally, by considering the interconnections between intra-day loads, cDDPMs establish a probabilistic distribution for the entire day to  address intra-day dependency issues. Furthermore, since this approach generates curves that share the same distribution as the target day rather than directly predicting values, it eliminates forecasting lag concerns. Recent work~\cite{wang2024diffload} proposed a Seq2Seq diffusion-based approach for probabilistic load forecasting to quantify both epistemic and aleatoric uncertainty. The method assumes that aleatoric uncertainty in load forecasting follows a Cauchy distribution. While this heavy-tailed distribution provides robustness against outliers, it is a fixed assumption and may not always align with real-world data distributions, which could exhibit mixed or non-parametric characteristics.

In this paper, rather than imposing strict parametric distribution assumptions, we propose a diffusion-based generative framework \modelname,  which  integrates neural networks such as LSTM and our developed TMSAB mechanism to learn the underlying distribution of load data directly.  The primary contributions of our approach are summarized below:

\begin{enumerate}
  \item The paper introduces a novel diffusion-based framework  tailored for PDALF. By progressively noising and denoising time-series data in a Markov chain, the model can effectively learn the underlying distribution of load curves and generate forecasts in a probabilistic manner.
  \item We design a new denoising network, \modelname,  specifically for load data, in which we develop an original attention mechanism called TMSAB. This mechanism simultaneously considers positional and temporal information within the sequence to more accurately capture the correlations between load data.
  \item We reconstruct the distribution of the real and generated load curves using KDE and compare them using KL divergence. Experimental results show that the proposed model, \modelname, can effectively fit the original data distribution and outperforms other benchmarks.
\end{enumerate}

The rest of this paper is structured as follows: Section \uppercase\expandafter{\romannumeral2} covers the concepts of cDDPM. Section \uppercase\expandafter{\romannumeral3} provides the technical details of the denoising network, \modelname. Section \uppercase\expandafter{\romannumeral4} presents the case studies. Finally, Section \uppercase\expandafter{\romannumeral5} concludes the paper.

\section{Diffusion-based Forecasting Model}

\begin{figure*}[t] 
    \centering
    \includegraphics[width=\textwidth]{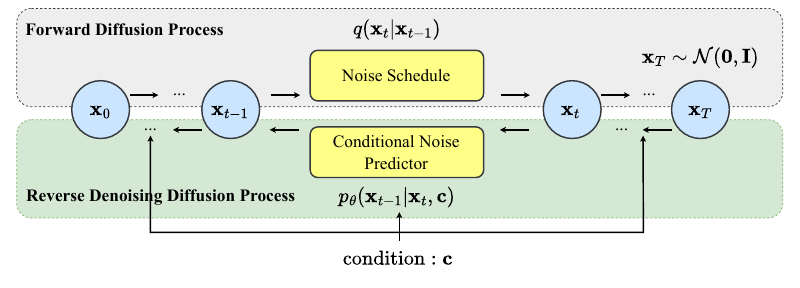} 
    \captionsetup{justification=raggedright, singlelinecheck=false, font=small} 
    \caption{Overview of the conditional diffusion probabilistic models.}
    \label{fig: overview od cDDPM}
\end{figure*}

In this section, we explain the principles of denoising diffusion models. The overview of the cDDPM is illustrated in Fig.~\ref{fig: overview od cDDPM}. The original DDPM is an unconditional generative model that learns a distribution $p_{\theta}(\mathbf{x}_{0})$ to approximate the true data distribution $q(\mathbf{x}_0)$. It consists of two processes: the forward process and the reverse process. In the forward process, Gaussian noise is progressively added to the original data sample $\mathbf{x}_{0}$, producing increasingly noisy versions $\mathbf{x}_{1},...,\mathbf{x}_{T}$ until $\mathbf{x}_{T}$ becomes nearly pure Gaussian noise. The reverse process then learns to gradually denoise $\mathbf{x}_{T}$ step by step to reconstruct $\mathbf{x}_{0}$.
In our work, we employ the conditional extension (cDDPM), where generation is guided by external variables (e.g., weather and calendar features). This allows the model to learn $p_\theta(\mathbf{x}_{0}\mid \mathbf{c})$, which captures the distribution of load curves under given conditions.

\subsection{Forward Diffusion Process}
Diffusion models have gained significant attention for their ability to generate high-dimensional data with excellent training stability compared to other generative models. Our approach begins by applying forward diffusion to convert real load curves into noisy samples. We then use reverse diffusion to reconstruct the load time series from this perturbed noise. The forward and reverse processes are modeled as Markov chains, with the reverse Gaussian transitions learned via a deep denoising neural network. Gaussian noise is incrementally added to the original load time-series data $\mathbf{x}_0$ during the forward process. After $T$ diffusion steps, the original distribution, $q(\mathbf{x}_0)$ is transformed into a standard Gaussian distribution $q(\mathbf{x}_T)$. This process of adding noise can be described as a fixed Markov chain:
\begin{equation}
    q(\mathbf{x}_{1:T}|\mathbf{x}_0)=\prod_{t=1}^{T}q(\mathbf{x}_t|\mathbf{x}_{t-1}),
\end{equation}
where $\mathbf{x}_{1},...,\mathbf{x}_{T}$ can be viewed as latent variables representing the intermediate states that result from perturbing the actual load curves with Gaussian noise at each step $t$. The term $q(\mathbf{x}_t|\mathbf{x}_{t-1})$ represents the forward Markov transition, specifying the mean and variance of the Gaussian noise added to $\mathbf{x}_{t-1}$:
\begin{equation}\label{qxtxt-1}
    q(\mathbf{x}_t|\mathbf{x}_{t-1})=\mathcal{N}(\mathbf{x}_t;\sqrt{1-\beta_t}\mathbf{x}_{t-1},\beta_t\mathbf{I}),
\end{equation}
where $\{\beta_1,...,\beta_T\}$ is an increasing variance schedule with $\beta_t \in (0,1)$ that represents the noise level at forward step $t$. Unlike typical latent variable models such as the variational autoencoder (VAE), the approximate posterior distribution $q(\mathbf{x}_t|\mathbf{x}_{t-1})$ in diffusion models is not trainable but is fixed to follow the aforementioned Gaussian transition process. Following the $T$-step diffusion process, a certain load curve is ultimately transformed into a straightforward Gaussian noise $\mathbf{x}_T$, which facilitates easier sampling and manipulation.

Let $\alpha_t=\prod_{n=1}^{t}(1-\beta_n)$, a particular property of the forward process is that the distribution of $\mathbf{x}_t$ given $\mathbf{x}_0$ has a close form:
\begin{equation}\label{qxtx0}
    q(\mathbf{x}_t|\mathbf{x}_{0})=\mathcal{N}(\mathbf{x}_t;\sqrt{\alpha_t}\mathbf{x}_{0},(1-\alpha_t)\mathbf{I}).
\end{equation}

Using the reparameteriztioin trick and $\boldsymbol{\epsilon}\sim\mathcal{N}(\mathbf{0},\mathbf{I})$ as a sampled noise, \eqref{qxtx0} can be expressed as:
\begin{equation}\label{x_t}
    \mathbf{x}_t=\sqrt{\alpha_t}\mathbf{x}_0+\sqrt{1-\alpha_t}\boldsymbol{\epsilon}.
\end{equation}

Based on \eqref{x_t}, we can directly sample $\mathbf{x}_t$ at any arbitrary noise level $t$, instead of computing the forward process step by step. The detailed proof of \eqref{qxtx0} and \eqref{x_t} can be found in \cite{luo2022understanding}.

\subsection{Reverse Denoising Process}

In contrast to the noise addition in the forward process, the reverse process progressively removes noise from the initial standard Gaussian noise $\mathbf{x}_T$ to recover the original load curve $\mathbf{x}_0$. The reverse process also follows a Markov chain with learnable Gaussian transitions starting from $p(\mathbf{x}_T)=\mathcal{N}(\mathbf{x}_T;\mathbf{0},\mathbf{I})$. Inspired by non-equilibrium thermodynamics, if we know the step-wise Gaussian noise applied in the forward process, we can restore the real load curve distribution through a series of iterative denoising steps. Such reverse process can be represented as follows.
\begin{equation}\label{back joint probability}
    p_{\theta}(\mathbf{x}_{0:T})=p(\mathbf{x}_{T})\prod_{t=1}^{T}p_{\theta}(\mathbf{x}_{t-1}|\mathbf{x}_{t}),
\end{equation}
where $\mathbf{x}_{T}\sim\mathcal{N}(\mathbf{0},\mathbf{I})$ and $p_{\theta}(\cdot)$ represents the reverse Markov transition. Based on the statistical characteristics of the continuous diffusion process outlined in \cite{ho2020denoising}, when the added Gaussian noise is sufficiently small, the denoising transition $p_{\theta}(\mathbf{x}_{t-1}|\mathbf{x}_{t})$ will resemble the functional form of the forward transition $q(\mathbf{x}_t|\mathbf{x}_{t-1})$.
Therefore, $p_{\theta}$ can represent a learnable Gaussian transition, which can be approximated by a neural network with $\theta$ representing the network parameters. Moreover, the transition between two adjacent latent variables is indicated by:
\begin{equation}\label{backMarkov}
    p_{\theta}(\mathbf{x}_{t-1}|\mathbf{x}_{t})=\mathcal{N}(\mathbf{x}_{t-1};\mu_{\theta}(\mathbf{x}_t,t),\Sigma_{\theta}(\mathbf{x}_t,t)),
\end{equation}
with shared parameters $\theta$, Here, we adopt the same parameterization of $p_{\theta}(\mathbf{x}_{t-1}|\mathbf{x}_{t})$ as in \cite{ho2020denoising} due to its demonstrated effectiveness in image generation:
\begin{subequations}
\label{mean and variance}
\begin{align}
    \mu_{\theta}(\mathbf{x}_t,t) &= \frac{1}{\sqrt{1-\beta_t}} \left( \mathbf{x}_t - \frac{\beta_t}{\sqrt{1-\alpha_t}} \epsilon_{\theta}(\mathbf{x}_t,t) \right) \label{eq:first}\\
    \Sigma_{\theta}(\mathbf{x}_t,t) &= \frac{1-\alpha_{t-1}}{1-\alpha_t} \beta_t,
\end{align}
\end{subequations}
where $\epsilon_{\theta}(\cdot)$ is a trainable denoising function determining the amount of noise to remove at each denoising step. Our goal is to design an efficient method to learn how to sample from $p_{\theta}(\mathbf{x}_{t-1}|\mathbf{x}_{t})$ and ultimately derive the load curve distribution $p_{\theta}(\mathbf{x}_0)$.

\subsection{Training Objective}
Instead of optimizing \eqref{backMarkov} directly due to its complexity, we approximate the real distribution of load curves by maximizing their log-likelihoods. Furthermore, directly solving the log$p_{\theta}(\mathbf{x}_{0})$ is impractical and difficult to compute explicitly. Hence, we typically opt to optimize its Evidence Lower Bound (ELBO) instead, which can be described as follows:
\begin{equation}
\label{ELBO Derive}
\begin{split}
    {\rm{log}}p_{\theta}(\mathbf{x}_0) &= {\rm{log}}\int p_{\theta}(\mathbf{x}_{0:T})d(\mathbf{x}_{1:T}) \\ &= {\rm{log}}\int \frac{p_{\theta}(\mathbf{x}_{0:T})}{q(\mathbf{x}_{1:T}|\mathbf{\mathbf{x}}_0)}q(\mathbf{x}_{1:T}|\mathbf{\mathbf{x}}_0)d(\mathbf{x}_{1:T}) \\ &= {\rm{log}}(\mathbb{E}_{q(\mathbf{x}_{1:T}|\mathbf{\mathbf{x}}_0)}[\frac{p_{\theta}(\mathbf{x}_{0:T})}{q(\mathbf{x}_{1:T}|\mathbf{\mathbf{x}}_0)}]) \\ &\geq \mathbb{E}_{q(\mathbf{x}_{1:T}|\mathbf{\mathbf{x}}_0)}[{\rm{log}}\frac{p_{\theta}(\mathbf{x}_{0:T})}{q(\mathbf{x}_{1:T}|\mathbf{\mathbf{x}}_0)}].
\end{split}
\end{equation}

The final inequality in \eqref{ELBO} is derived from Jensen’s inequality. It is important to note that $q(\mathbf{x}_{1:T}|\mathbf{\mathbf{x}}_0)$ is the posterior distribution of $p_{\theta}(\mathbf{x}_{0})$, defined by a sequence of latent variables $\mathbf{x}_{1},...,\mathbf{x}_{T}$ in the forward process, and it can be readily computed using \eqref{qxtxt-1}. Consequently, maximizing ${\rm{log}}p_{\theta}(\mathbf{x}_{0:T})$ is equivalent to minimizing its negative ELBO. This negative ELBO can be decomposed into $T+1$ tractable items:
\begin{equation}\label{ELBO}
    ELBO=-(\mathcal{L}_{0}+\Sigma_{T}^{t=2}\mathcal{L}_{t-1}+\mathcal{L}_T),
\end{equation}
the exact expressions of $\mathcal{L}_{0}$, $\mathcal{L}_{t-1}$, and $\mathcal{L}_{T}$ are illustrated as follows: 
\begin{subequations}
\label{-ELBO}
\begin{align}
    \quad & \mathcal{L}_{0} = -\mathbb{E}_{q(\mathbf{x}_{1}|\mathbf{\mathbf{x}}_0)}[{\rm{log}}p_{\theta}(\mathbf{x}_0|\mathbf{x}_1)] \\
    \quad & \mathcal{L}_{t-1}= \mathbb{E}_{q(\mathbf{x}_{t}|\mathbf{\mathbf{x}}_0)}[\mathcal{D}_{KL}(q(\mathbf{x}_{t-1}|\mathbf{x}_t,\mathbf{x}_0)||p_{\theta}(\mathbf{x}_{t-1}|\mathbf{x}_t))]\\
    \quad & \mathcal{L}_{T}=\mathcal{D}_{KL}(q(\mathbf{x}_T|\mathbf{x}_0)||p_{\theta}(\mathbf{x}_T)),
\end{align}
\end{subequations}
where $\mathcal{D}_{KL}$ is the KL Divergence, and $\mathcal{L}_{0},\mathcal{L}_{t-1},\mathcal{L}_{T}$ are referred to as the reconstruction term, the consistency term, and the prior matching term, respectively. The term $\mathcal{L}_{0}$ predicts the log probability of the original data sample given the first-step latent. Additionally, $\mathcal{L}_{0}$ is a special case of $\mathcal{L}_{t-1}$. When $t=1$, $\mathcal{L}_{t-1}$ transforms into $\mathcal{L}_{0}$. $\mathcal{L}_{t-1}$ aims to ensure consistency in the distribution at $\mathbf{x}_t$ for both the forward and backward processes, which means that each denoising step from a noisier image should correspond to the appropriate noising step from a cleaner image at every intermediate timestep. 
The term $\mathcal{L}_{T}$ does not require optimization since it has no trainable parameters. Moreover, KL Divergence is used to measure the difference between two distributions. The smaller the difference, the smaller the value. When the two distributions are identical, the KL Divergence is 0. Given our assumption of a sufficiently large $T$ such that the final distribution is Gaussian, this term effectively becomes zero. The KL Divergence mathematically represents this relationship. The term $q(\mathbf{x}_{t-1}|\mathbf{x}_t,\mathbf{x}_0)$ can be analytically calculated as follows:
\begin{equation}
\label{bayesian}
\begin{split}
    q(\mathbf{x}_{t-1}|\mathbf{x}_t,\mathbf{x}_0)&=\frac{q(\mathbf{x}_{t}|\mathbf{x}_{t-1},\mathbf{x}_0)q(\mathbf{x}_{t-1}|\mathbf{x}_0)}{q(\mathbf{x}_{t}|\mathbf{x}_0)}\\ &= \mathcal{N}(\mathbf{x}_{t-1};\widetilde{\mu}_t(\mathbf{x}_{t},\mathbf{x}_{0}),\widetilde{\beta}_t\mathbf{I}).
\end{split}
\end{equation}

The first line of \eqref{bayesian} utilizes the Bayesian rule, 
 which makes the conditional probability tractable since the right-hand side of \eqref{bayesian} involves three known Gaussian distributions. The mean and variance defined in \eqref{bayesian} are as follows:
\begin{equation}
\label{mean bayesian}
\begin{split}
    \widetilde{\mu}_t(\mathbf{x}_{t},\mathbf{x}_{0}) &=\frac{\sqrt{\alpha_{t-1}}\beta_{t}}{1-\alpha_t}\mathbf{x}_0+\frac{\sqrt{1-\beta_{t-1}}(1-\alpha_{t-1})}{1-\alpha_t}\mathbf{x}_t \\ &=  \frac{1}{\sqrt{1-\beta_t}}(\mathbf{x}_t-\frac{\beta_t}{\sqrt{1-\alpha_t}}\boldsymbol{\epsilon}),
\end{split}
\end{equation}

\begin{equation}\label{variance bayesian}
    \widetilde{\beta}_t=\frac{1-\alpha_{t-1}}{1-\alpha_t}\beta_t.
\end{equation}
In \eqref{bayesian}, it should be noted that $\mathbf{x}_0$ can be derived from $\mathbf{x}_t$ and $\boldsymbol{\epsilon}$ based on \eqref{x_t}. Following the empirical simplification adopted in \cite{ho2020denoising}, we also fix $\Sigma_{\theta}(\mathbf{x}_t,t)=\widetilde{\beta}_t\mathbf{I}$ for $p_{\theta}(\mathbf{x}_{t-1}|\mathbf{x}_t)$, which eliminates the need to learn the variance of the reverse transition. Consequently, $\mathcal{L}_{t-1}$ reduces to calculating the KL divergence between the two Gaussian distributions defined in \eqref{backMarkov} and \eqref{bayesian}, which have different means but the same variance. Therefore, $\mathcal{L}_{t-1}$ can be computed as follows:
\begin{equation}\label{Lt-11}
    \mathcal{L}_{t-1}=\mathbb{E}_{q(\mathbf{x}_T|\mathbf{x}_0)}[\frac{1}{2\widetilde{\beta}_t}||\widetilde{\mu}_t(\mathbf{x}_t,\mathbf{x}_0)-\mu_{\theta}(\mathbf{x}_t,t)||_{2}^{2}],
\end{equation}
which aims to predict the mean of the reverse transition. According to \eqref{mean bayesian} and the method proposed in \cite{ho2020denoising}, an alternative parameterization for $\mu_{\theta}(\mathbf{x}_t,t)$ can be defined in \eqref{eq:first}. This parameterization suggests a more efficient way for training the denoising network, specifically by directly predicting the Gaussian noise $\boldsymbol{\epsilon}$ added at step $t$ instead of the mean $\widetilde{\mu}_t$ of the reverse transition. Consequently, $\mathcal{L}_{t-1}$ can be reformulated as follows:
\begin{equation}\label{Lt1}
    \mathcal{L}_{t-1}=\mathbb{E}_{\mathbf{x}_0,\boldsymbol{\epsilon},t}[||\boldsymbol{\epsilon}-\epsilon_{\theta}(\sqrt{\alpha_t}\mathbf{x}_0+\sqrt{1-\alpha_t}\boldsymbol{\epsilon},t)||_{2}^{2}],
\end{equation}
where $t$ is uniformly sampled from $[1,T]$, and $\epsilon_{\theta}(\cdot)$ denotes the denoising neural network. $\epsilon_{\theta}(\cdot)$ takes the perturbed $\mathbf{x}_t$ defined in \eqref{x_t} as input and outputs the Gaussian noise prediction $\hat{\boldsymbol{\epsilon}}$ at step $t$.  After training the denoising network $\epsilon_{\theta}(\cdot)$ using \eqref{Lt1}, it can be used to incrementally reconstruct the load curves during the reverse denoising process by computing $\mu_{\theta}(\mathbf{x}_t,t)$ as given in \eqref{eq:first}.

\subsection{Conditional Diffusion Model}
The standard DDPM is intended to generate images from pure white noise without any conditions, which is not suitable for our task of generating future load curves based on historical data. To address this, we introduce cDDPM, where the model is conditioned on past load information to produce the desired future load curves.

In the original DDPM, the reverse process $ p_{\theta}(\mathbf{x}_{0:T})$ as outlined in \eqref{back joint probability} is employed to estimate the final data distribution $ q(\mathbf{x}_{0})$. For the cDDPM, this equation needs to be adjusted as follows:
\begin{equation}\label{conditional DDPMs joint}
    p_{\theta}(\mathbf{x}_{0:T}|\mathbf{c})=p(\mathbf{x}_{T})\prod_{t=1}^{T}p_{\theta}(\mathbf{x}_{t-1}|\mathbf{x}_{t},\mathbf{c}),
\end{equation}
where $\mathbf{c}$ represents the condition. Similarly, \eqref{backMarkov} and \eqref{Lt1} are revised as:

\begin{equation}\label{conditional backMarkov}
    p_{\theta}(\mathbf{x}_{t-1}|\mathbf{x}_{t},\mathbf{c})=\mathcal{N}(\mathbf{x}_{t-1};\mu_{\theta}(\mathbf{x}_t,\mathbf{c},t),\widetilde{\beta}_t\mathbf{I})),
\end{equation}
\begin{equation}\label{conditional Lt1}
    \mathcal{L}_{t-1}^{cond}=\mathbb{E}_{\mathbf{x}_0,\boldsymbol{\epsilon},t}[||\boldsymbol{\epsilon}-\epsilon_{\theta}(\sqrt{\alpha_t}\mathbf{x}_0+\sqrt{1-\alpha_t}\boldsymbol{\epsilon},\mathbf{c},t)||_{2}^{2}].
\end{equation}

To maximize the ELBO, we need to minimize $\mathcal{L}_{t-1}^{cond}$. Therefore, the optimization objective can be defined as follows:
\begin{equation}\label{optimization objective}
    \min_{\theta}\mathcal{L}{(\theta)}=\min_{\theta}\mathbb{E}_{\mathbf{x}_0,\boldsymbol{\epsilon},t}[||\boldsymbol{\epsilon}-\epsilon_{\theta}(\sqrt{\alpha_t}\mathbf{x}_0+\sqrt{1-\alpha_t}\boldsymbol{\epsilon},\mathbf{c},t)||_{2}^{2}].
\end{equation}

According to \eqref{optimization objective}, we ultimately need to design a function $\epsilon_{\theta}(\cdot)$ to predict the noise. $\epsilon_{\theta}(\cdot)$ takes the original sample $\mathbf{x}_{0}$, the condition $\mathbf{c}$, standard Gaussian noise $\boldsymbol{\epsilon}$, and the noise addition step $t$ as inputs. The expectation calculation $\mathbb{E}$ can be accomplished through uniform sampling. \eqref{optimization objective} represents the final implementation of cDDPM, and the training procedure  is presented in Algorithm 1.

\begin{algorithm}[!h]\label{alg:training}
  \caption{Training of cDDPMs} 
  \begin{algorithmic}[1]
  \Require Load curve datasets with conditions $q(\mathbf{x}_0|\mathbf{c})$.
  \Ensure Denoising diffusion model $\epsilon_{\theta}(\cdot)$.
  \While{$\theta$ has not converged}
        \State $\mathbf{x}_0\sim q(\mathbf{x}_0|\mathbf{c})$, $\boldsymbol{\epsilon}\sim \mathcal{N}(\mathbf{0},\mathbf{I}), t\sim$ Uniform(\{${1,...,T}\}$)
        \State Update the denoising network using gradient descent: $\nabla||\boldsymbol{\epsilon}-\epsilon_{\theta}(\sqrt{\alpha_t}\mathbf{x}_0+\sqrt{1-\alpha_t}\boldsymbol{\epsilon},\mathbf{c},t||_{2}^{2}$ using \eqref{conditional Lt1}
  \EndWhile
  \end{algorithmic}
\end{algorithm}

Once a noise-predicting cDDPM is trained, we can restore the original curves by progressively removing the noise through reverse. The algorithm for the reverse process is illustrated in Algorithm 2.

\begin{algorithm}[!h]\label{alg:sampling}
  \caption{Sampling of cDDPMs} 
  \begin{algorithmic}[1]
  \Require Gaussian noise $\mathbf{x}_T\sim\mathcal{N}(\mathbf{0},\mathbf{I})$, condition $\mathbf{c}$.
  \Ensure Generated load curves.
    \For{$t=T,...,1$}
        \State $\mathbf{z}\sim\mathcal{N}(\mathbf{0},\mathbf{I})$
        \State $\mathbf{x}_{t-1}=\frac{1}{\sqrt{1-\beta_t}}(\mathbf{x}_t-\frac{\beta_t}{\sqrt{1-\alpha_t}}\epsilon_{\theta}(\mathbf{x}_t,\mathbf{c},t))+\sqrt{\widetilde{\beta}_t}\mathbf{z}$ according to both \eqref{backMarkov} and \eqref{eq:first}.
    \EndFor
  \end{algorithmic}
\end{algorithm}

\section{Details of Denoising Architecture}

According to \eqref{optimization objective}, we need to design a denoising network $\epsilon_{\theta}(\cdot)$ to predict the noise and progressively remove the predicted noise to generate the day-ahead load curve. Therefore,  designing $\epsilon_{\theta}(\cdot)$ that inputs historical load data is crucial for constructing an effective diffusion model. This section will illustrate how to design a denoising network \modelname\ suitable for load time series.

\begin{figure}[h]
  \centering
  \includegraphics[width=4.49in]{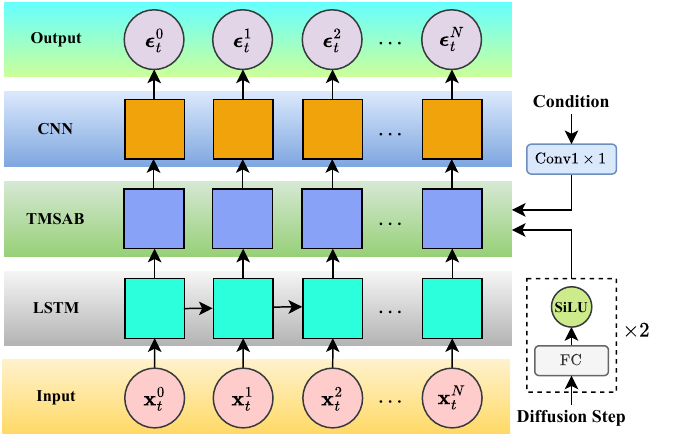}
  \captionsetup{justification=raggedright, singlelinecheck=false, font=small} 
  \caption{Structure of the proposed denoising network DALNet.}
  \label{fig: Overview of Denoising Network}
\end{figure}

\subsection{Overview of Denoising Network}
Fig. \ref{fig: Overview of Denoising Network} depicts the overall structure of the denoising model \modelname. Given that our load data is a time series with distinct temporal characteristics, it is intuitive to design a neural network capable of extracting these features as the primary framework for $\epsilon_{\theta}(\cdot)$. This architecture aims to preserve the temporal associations during the reverse generation process. Therefore, we utilized LSTM, a classic neural network for processing time series, and our improved attention mechanisms, TMSAB, which have demonstrated advantages in handling textual data. Given an input sample $\mathbf{x}_{0}$, we can obtain the noise sample $\mathbf{x}_{t} \in \mathbb{R}^{N \times 1}$ using \eqref{x_t}, where $N$ presents the length of time-series $\mathbf{x}_{t}$. The output of LSTM can be expressed as $\mathbf{h}_{t}=LSTM(\mathbf{x}_{t})$, with $\mathbf{h}_{t}\in \mathbb{R}^{N \times H}$. $H$ is the dimension of the hidden state vector. Another reason for choosing LSTM is that it effectively alleviates the vanishing gradient problem and offers enhanced nonlinearity due to its multiple gate functions. Details about LSTM can be found in \cite{LSTM}. The output of the LSTM, combined with the diffusion step embedding and condition embedding, serves as the input for the TMSAB, which not only captures temporal correlations in the load curve but also extracts positional information within the sequence. The output of the attention mechanism is fed into 1D-convolutional layers, which generate the final noise prediction result $\boldsymbol{\epsilon}_{t} \in \mathbb{R}^{N \times 1}$.

\subsection{Temporal Multi-scale Attention Mechanism}

In \cite{vaswani2017attention}, a multi-head attention mechanism is introduced. This approach computes attention scores through linear transformations applied to the queries, keys, and values $(Q, K, V)$. Assuming the dimension of the multi-head attention input $\mathbf{x} = \{x_1,...,x_n,...,x_N\}$ is $\mathbb{R}^{N\times H}$, where $N$ represents the sequence length and the dimension of $x_n$ is $1\times H$. First, by focusing on one of the heads, the attention mechanism applies a linear transformation to each element in the sequence to generate $Q=\{q_1,...,q_n,...,q_N\}$ and $K=\{k_1,...,k_n,...,k_N\}$: $Q=\mathbf{x}*\mathbf{w}_Q, K=\mathbf{x}*\mathbf{w}_K$. Both $Q$ and $K$ have the same dimension of $\mathbb{R}^{N\times M}$, and $M$ denotes the dimension of the attention model, which means the element in the sequence has its corresponding $q$ and $k$. By multiplying the $q_i$ and $k_j$, we can derive an attention score $a_{ij}$, representing the relationship between the $i$-th element and the $j$-th element in the sequence. In a similar way, by multiplying $Q$ and $K$, we can obtain an attention map $\mathbf{A}$ with the dimension of $\mathbb{R}^{N\times N}$.
\begin{equation}\label{attention map}
    \mathbf{A} = \mathrm{softmax}{(\frac{QK^T}{\sqrt{M}})}.
\end{equation}

\begin{figure}[h]
  \centering
  \includegraphics[width=5.49in]{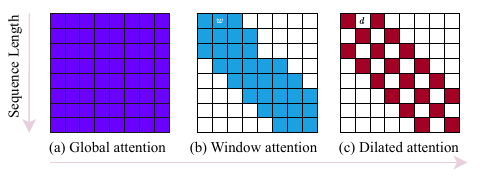}
  \captionsetup{justification=raggedright, singlelinecheck=false, font=small} 
  \caption{Illustration of three attention maps. (a): Global attention map; (b): Window attention map with window size $w=2$; (c) Dilated window attention map with dilation $d=1$.}
  \label{figure Attention map}
\end{figure}

The diagonal elements of the attention map indicate the self-attention scores, while the other elements represent the mutual attention scores between different elements. This type of attention mechanism is also called global attention, with the diagram of its attention map shown in Fig. \ref{figure Attention map}(a). When considering multiple heads, we obtain multiple attention maps.

After generating the attention map, it can be transformed into the output $O$ of one head through a linear transformation matrix $V$: $V=\mathbf{x}*\mathbf{w}_V$ and $O = \mathrm{softmax}{(\frac{QK^T}{\sqrt{M}})}V$. For multi-head attention, we stack outputs of all heads and also use a linear transformation:
\begin{equation}\label{multi-attention}
    \mathrm{Multihead} = [O_1,O_2,...O_h]*\mathbf{w}_O,
\end{equation}
where $h$ is the number of the attention heads.

A critical factor in its effectiveness is that the self-attention component enables the model to grasp contextual information across the entire sequence. Despite its power, the self-attention mechanism has significant drawbacks: its memory and computational requirements escalate quadratically with the length of the sequence, making it inefficient or prohibitively expensive to handle long sequences. What is more, Algorithm 2 reveals that the sampling process in the diffusion model requires iteration from $T$ down to 1. When $T$ is large and the sequence is lengthy, using a global attention mechanism significantly increases the computation time needed. Besides the computational challenge, \cite{improvingattention} highlights that the diversity of multi-head attention needs to be improved. Often, the attention heads are highly repetitive and need to focus on distinct representation subspaces as intended.

Therefore, to address the issues above, we introduced the dilated window attention mechanism as described in \cite{longformer}. According to \eqref{attention map}, we know that for a sequence $\mathbf{x}$ each of its elements computes an attention score $a_{ij}$ with all other elements, while window attention refers to the mechanism where, for the $i$-th element in a sequence $\mathbf{x}$, the attention score is calculated only with a few neighboring elements. A diagram of the window attention map can be found in Fig. \ref{figure Attention map}(b). In this way, it reduces the computational cost and places greater emphasis on the weights of the elements adjacent to the $i$-th element. In addition, to further increase the diversity of the attention map, we also incorporated dilated window attention, which can be found in Fig. \ref{figure Attention map}(c).

\begin{figure}[h]
  \centering
  \includegraphics[width=5.49in]{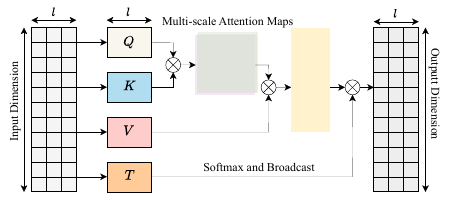}
  \captionsetup{justification=raggedright, singlelinecheck=false, font=small} 
  \caption{Visualization of the TMSAB with the sequence length $l$.}
  \label{fig: visualization of TMSAB}
\end{figure}

Although the multi-scale attention map addresses the issue of highly repetitive attention heads, it tends to focus more on positional information within the sequence while neglecting temporal information \cite{TemporalAttention}. To resolve this issue, we introduced a temporal embedding $T$ and  proposed a new attention mechanism called TMSAB, with the visualization of TMSAB illustrated in Fig. \ref{fig: visualization of TMSAB}. Like the method of generating $Q, K, V$ in global attention, we first apply a linear transformation $T=\mathbf{x}*\mathbf{w}_T$ to the input data to convert it into $T$. Then, we use an activation function $\mathrm{tanh(\cdot)}$ to mitigate the vanishing gradient problem, followed by a transformation into an $l\times1$, where $l$ is the sequence length. Finally, a temporal attention vector is generated using $\mathrm{softmax}$, and broadcasting ensures it meets the required dimensions. The complete expression for the temporal attention $\mathbf{T}$ is as follows:
\begin{equation}\label{eq: Temporal embedding}
    \mathbf{T} = \mathrm{softmax}(\mathrm{tanh}(\mathbf{x}*\mathbf{w}_T)*\mathbf{w}_l).
\end{equation}

By designing TMSAB, we address the issue of highly repetitive attention maps in global attention and resolve the problem of multi-scale attention focusing solely on positional relationships. Therefore, introducing TMSAB is well-suited for load data, which is inherently a time series.

\subsection{Diffusion Step and Conditionality Embedding}
According to Algorithm 1 and 2, the denoising diffusion model $\epsilon_{\theta}(\cdot)$ has three inputs: the sample $\mathbf{x}_t$, the diffusion step $t$, and the condition $\mathbf{c}$. Notably, $t$ plays a crucial role in $\epsilon_{\theta}(\cdot)$ as different values of $t$ result in varying noise intensities. To incorporate the time step $t$ into the model, which is akin to adding positional information, position embedding as introduced by \cite{vaswani2017attention} has been employed, which involves converting $t$ into a vector comprising sine and cosine functions at different frequencies. A vector dimension of 64 is chosen for embedding the time step $t$, which can be represented as follows:
\begin{equation}\label{time embedding}
\begin{split}
    t_{\text{embedding}} = [&\sin\left(10^{\frac{0 \times 4}{31}} t\right), \ldots, \sin\left(10^{\frac{31 \times 4}{31}} t\right), \\
                            &\cos\left(10^{\frac{0 \times 4}{31}} t\right), \ldots, \cos\left(10^{\frac{31 \times 4}{31}} t\right)].
\end{split} 
\end{equation}

Once the embedding is generated, it passes through two fully connected layers, each utilizing a SiLU activate function: $\mathrm{SiLU}(\mathrm{FC}(\mathrm{SiLU}(\mathrm{FC}(t_{embedding}))))$. The choice of SiLU is due to its ability to alleviate the vanishing gradient problem \cite{SiLU}.

Theoretically, for condition embedding, day-ahead load generation can utilize various conditions, such as the predicted highest and lowest loads for the day ahead, and other meteorological forecast data like temperature and humidity. However, we opt not to use these conditions because they require prior predictions, and any inaccuracies in the load and weather forecasts would lead to inaccuracies in the generated load curve and results in cumulative errors. We use the previous day's load as the condition to avoid using forecasted values inspired by \cite{DiffSTG}. Consequently, the dimensions of the condition should be the same as those of the generated load curve. This condition vector $\mathbf{c}$ is processed through multiple $1\times 1$ convolutions to meet the dimensional requirements.

\subsection{Noise Schedule}
According to Algorithm 2, the Gaussian noise subtracted during the sampling process varies in intensity, with larger values of $\frac{\beta_t}{\sqrt{1-\alpha_t}}$ indicating more substantial subtracted noise. The selection of $\beta_t$ is crucial for the denoising model's ability to generate accurate load curves. Based on \cite{nichol2021improved}, we employ a quadratic schedule because it allows for smoothly adding noise incrementally, which enhances the generative ability of the diffusion model:
\begin{equation}\label{noise schedule}
    \beta_t = (\frac{T-t}{T-1}\sqrt{\beta_1}+\frac{t-1}{T-1}\sqrt{\beta_T})^2.
\end{equation}

We follow the standard setting: $\beta_1=0.0001$ and $\beta_T = 0.5$. Based on this noise schedule, the noise coefficients $\sqrt{\alpha_t}$ and $\sqrt{1-\alpha_t}$ in \eqref{x_t} can be calculated. As $t$ increases, $\sqrt{\alpha_t}$ gradually decreases while $\sqrt{1-\alpha_t}$ increases. According to \eqref{x_t}, $\sqrt{\alpha_t}$ is the coefficient for the original sample $\mathbf{x}_0$, and $\sqrt{1-\alpha_t}$ is the coefficient for the noise $\boldsymbol{\epsilon}$, indicating that as $t$ increases, the original sample progressively transforms into a noisy sample, with the degree of noise increasing. Furthermore, the original sample completely transforms into Gaussian noise when $t=T$, $\sqrt{\alpha_t}=0$. This perspective also indicates that the quadratic noise schedule is reasonable.

\section{Case Study}
This section will elaborate on the dataset, evaluation metrics, benchmarks, and experimental results.

\subsection{Data and Evaluation Criteria Description}
We utilized the GEFcom2014 dataset, which includes four subsets: load, wind power, solar power, and electricity price. For our study, we focused on the load data to evaluate the performance of the proposed diffusion model. This load dataset spans 3,560 days, covering the period from January 1, 2001, at 1:00 AM to October 1, 2010, at 00:00, with a data resolution of one hour. More detailed information about the data can be found in \cite{dataset}. The dataset was divided into training and testing sets with a train-test split ratio of 8:2 to ensure a robust evaluation.

To evaluate the proposed model and other benchmarks, we employed Reliability, Sharpness, and Overall Score as the evaluation metrics in this paper. Reliability is the primary metric for evaluating the quality of probabilistic forecasting models. Reliability implies that, given a specified prediction interval nominal confidence (PINC) $100(1-\alpha)\%$, where $\alpha$ is the significance level, the prediction interval coverage probability (PICP) should be as close to the PINC as possible. PINC generally represents the theoretical probability of the predicted intervals (PIs). At the same time, PICP is the probability that the actual load values fall within the PIs, with their difference called Absolute Coverage Error (ACE), where a smaller absolute value of ACE indicates more accurate predictions.

Sharpness refers to the extent to which the predicted distribution closely matches the actual distribution. It is measured by the average width (AW) of the PIs, which can be determined by generating intervals for all samples and calculating their average.

The final metric is the overall score, which can be calculated as follows:
\begin{subequations}
\label{equation: overall score}
\begin{align}
    S_t^{(\alpha)} &=
    \begin{cases}
        -2\alpha W_t^{(\alpha)}-4(\hat{L}_t^{(\alpha)}-y_i), & \text{if } y_i<\hat{L}_t^{(\alpha)} \\
        -2\alpha W_t^{(\alpha)}, & \text{if } y_i\in \hat{I}_t^{(\alpha)} \\
        -2\alpha W_t^{(\alpha)}-4(y_i-\hat{U}_t^{(\alpha)}), & \text{if } y_i>\hat{U}_t^{(\alpha)}
    \end{cases} \label{eq: S}\\
    \overline{S}_t^{(\alpha)} &= \frac{1}{|N|}\sum_{t\in N} S_t^{(\alpha)}, \label{eq: average S}
\end{align}
\end{subequations}
where $W$ represents the width of the PI at time step $t$, $I$ denotes the PI, and $L$ and $U$ represent the lower and upper bounds of the PI, respectively. \eqref{eq: S} indicates that for a given time $t$ if the actual load value falls within the PI, the absolute value of $S$ should be minimal. If not, an additional penalty term is added. \eqref{eq: average S} calculates the overall score by averaging the $S$ across all PIs. According to \eqref{equation: overall score}, $S$ is negative, and the smaller the absolute value, the better the forecasting performance.

\begin{table}[h]
\centering
\caption{Training hyperparameters of the proposed model.}
\label{tab:training_settings}
\renewcommand{\arraystretch}{1.15}
\setlength{\tabcolsep}{8pt}
\begin{tabular}{l l}
\toprule
\textbf{Parameter} & \textbf{Value} \\
\midrule
Diffusion steps & 1000 \\
Optimizer & Adam \\
Learning rate & $5\times 10^{-4}$ \\
Batch size & 64 \\
Training epochs & 60 \\
Loss function & Mean squared error on noise prediction \\
Hidden dimension & 128 \\
Dropout rate & 0.3 \\
\bottomrule
\end{tabular}
\end{table}

\subsection{Benchmarks}

The proposed model is trained with the hyperparameter settings presented in Table~\ref{tab:training_settings}. 
For comparison, we further evaluate several representative baselines, including Multi-layer perceptron (MLP), 
LSTM, Transformer, and Bayesian neural network (BNN). MLP is chosen as the benchmark for modeling different hours. Specifically, for this benchmark, we trained 24 MLPs, one for each hour, with the size of hidden layers chosen from $\{32, 64, 128\}$. LSTM, Transformer, and BNN are all used for direct multi-step prediction, meaning they simultaneously predict the load values for the next 24 hours. We did not choose to use the rolling prediction method as a benchmark because rolling prediction is more commonly used for point forecasting. LSTM, as a classic time series prediction model, is chosen as a benchmark with the hidden size chosen from $\{48, 96, 192\}$.

Similarly, the Transformer, which has demonstrated powerful natural language sequence processing capabilities, is also widely used in time series prediction tasks and is therefore chosen as a benchmark, with the number of encoder and decoder layers set to 2. The three methods mentioned above all implement quantile regression by introducing the pinball loss function to achieve probabilistic forecasting. For the quantile crossing problem, we resolved it using the naive rearrangement method as described in \cite{QuantileCrossing}.

As a probabilistic model, BNN does not require quantile regression to achieve probabilistic forecasting. BNN introduces probability distributions to the neural network's weights, which allows different weights and biases to be sampled during each inference. By performing multiple samples, probabilistic forecasting can be achieved. Therefore, BNN is also chosen as a benchmark. We implemented the BNN using convolutional layers and fully connected layers, with the number of channels chosen from $\{32, 64\}$ and the kernel size chosen from $\{(3,1), (2,1)\}$, respectively.

\subsection{Results Analysis of Evaluation Criteria}

\begin{table}[h]
\centering
\renewcommand{\arraystretch}{1.3}
\setlength{\tabcolsep}{7pt}
\caption{The Evaluation Criteria Results For Different Models}
\label{table: Evaluation Criteria}

\begin{tabular}{@{}c|l|c|c|c}
\toprule
\multicolumn{1}{c}{\textbf{PINC}} & \multicolumn{1}{c}{\textbf{Model}} & \multicolumn{1}{c}{\textbf{ACE}} & \multicolumn{1}{c}{\textbf{AW}} & \multicolumn{1}{c}{\textbf{\(\overline{\mathbf {S}}^{(\boldsymbol{\alpha})}\)}} \\
\midrule
\multirow{3}{*}{80\%}    & MLP                           & 2.72\% & 0.2110 & -0.1112 \\
                         & LSTM                          & 5.72\% & 0.1759 & -0.0898 \\
                         & Transformer                   & 4.95\% & 0.2154 & -0.1094 \\
                         & BNN                           & 4.26\% & 0.1814 & -0.0950 \\                   &\textbf{DALNet}  & 2.04\% & 0.1534 & -0.0682 \\
\midrule            
\multirow{3}{*}{90\%}    & MLP                           & 0.62\% & 0.2640 & -0.0657 \\
                         & LSTM                          & 3.06\% & 0.2265 & -0.0538 \\
                         & Transformer                   & 2.42\% & 0.2731 & -0.0650 \\
                         & BNN                           & 2.34\% & 0.2322 & -0.0541 \\                          
                         & \textbf{DALNet} & 2.01\% & 0.1941 & -0.0449 \\
\midrule
\multirow{3}{*}{95\%}    & MLP                           & 0.72\% & 0.3034 & -0.0376 \\
                         & LSTM                          & 1.16\% & 0.2671 & -0.0309 \\
                         & Transformer                   & 0.53\% & 0.3137 & -0.0370 \\
                         & BNN                           & 1.70\% & 0.2798 & -0.0354 \\                         
                         & \textbf{DALNet} & 1.15\% & 0.2172 & -0.0277 \\
\bottomrule
\end{tabular}
\end{table}

Table \ref{table: Evaluation Criteria} presents the ACE, AW, and Overall Score $\overline{S}^{(\alpha)}$ results of five different models at three different PINC. It can be observed that at 80\% PINC, the results of the proposed model are better than all other benchmarks. The ACE result is 0.68\% better than the second-best result (MLP). The AW and $\overline{S}^{(\alpha)}$ results are superior to LSTM by 0.0225 and 0.0216, respectively, which represent the second-best results. Similar to the results obtained at 80\% PINC, when the PINC is set at 90\%, the proposed model generally outperforms the benchmarks across the three metrics, except for the ACE, where it lags behind MLP by 1.39\%. LSTM continues to achieve the second-best results in both AW and $\overline{S}^{(\alpha)}$, trailing the proposed model by 0.0381 and 0.0089, respectively. For the PINC of 95\%, the proposed model ranks third in the ACE metric, with Transformer and MLP outperforming the proposed model by 0.62\% and 0.43\%, respectively. These results are mainly due to the significantly higher AW values obtained by these two models, which are greater than those of the proposed model by 0.0965 and 0.0862, respectively. Aside from these two models, LSTM continues to perform consistently, achieves the second-best results in AW and $\overline{S}^{(\alpha)}$, and trails the proposed model by 0.0499 and 0.0032, respectively.

Overall, the proposed model generally outperforms the benchmarks across the three metrics, particularly in AW and $\overline{S}^{(\alpha)}$, where it significantly surpasses other models. These results are because the diffusion model effectively captures the distribution of the original data and generates results that align well with the original data distribution.
For Transformer and BNN, part of their poorer performance is due to the difficulty in training these models. The Transformer requires a large number of parameters to be trained, and even when the number of encoder and decoder layers is set small, there are still over 100,000 parameters to train. As for BNN's training process, it involves Bayesian inference, which has a slow convergence rate and high computational complexity. Given the small sample size in our dataset, it is reasonable that the results for BNN could be more outstanding. 

\begin{figure}[!t]
  \centering
  \includegraphics[width=5.493in]{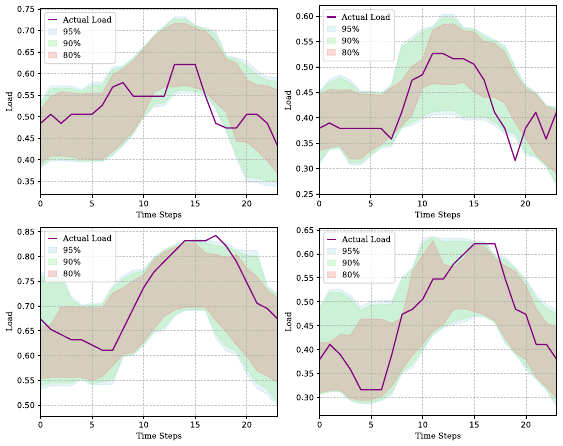}
  \captionsetup{justification=raggedright, singlelinecheck=false, font=small} 
  \caption{Probabilistic forecasting results of the DALNet for the three loads under three different
PINCs.}
  \label{figure: confidence intervals}
\end{figure}

Fig. \ref{figure: confidence intervals} presents the PIs generated by the diffusion models. It can be observed that the intervals produced by the diffusion model can capture the variation patterns of the intra-day load curves, as well as some minor fluctuations. Through Table \ref{table: Evaluation Criteria} and Fig. \ref{figure: confidence intervals}, it is evident that the proposed model shows considerable potential in PDALF tasks.

\begin{figure}[!t]
  \centering
  \includegraphics[width=4.52in]{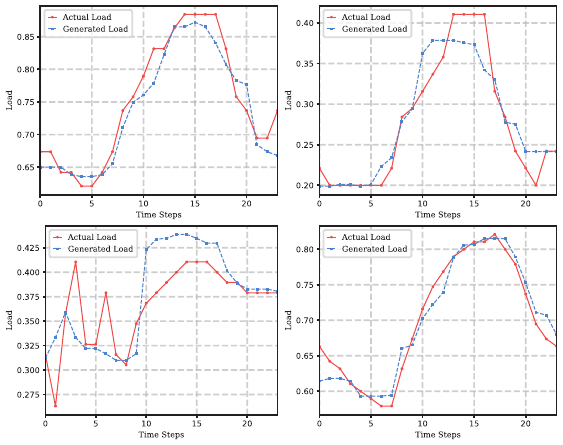}
  \captionsetup{justification=raggedright, singlelinecheck=false, font=small} 
  \caption{Point forecasting results of the DALNet.}
  \label{figure: load curves}
\end{figure}

\begin{figure*}[!t] 
    \centering
    \includegraphics[width=\textwidth]{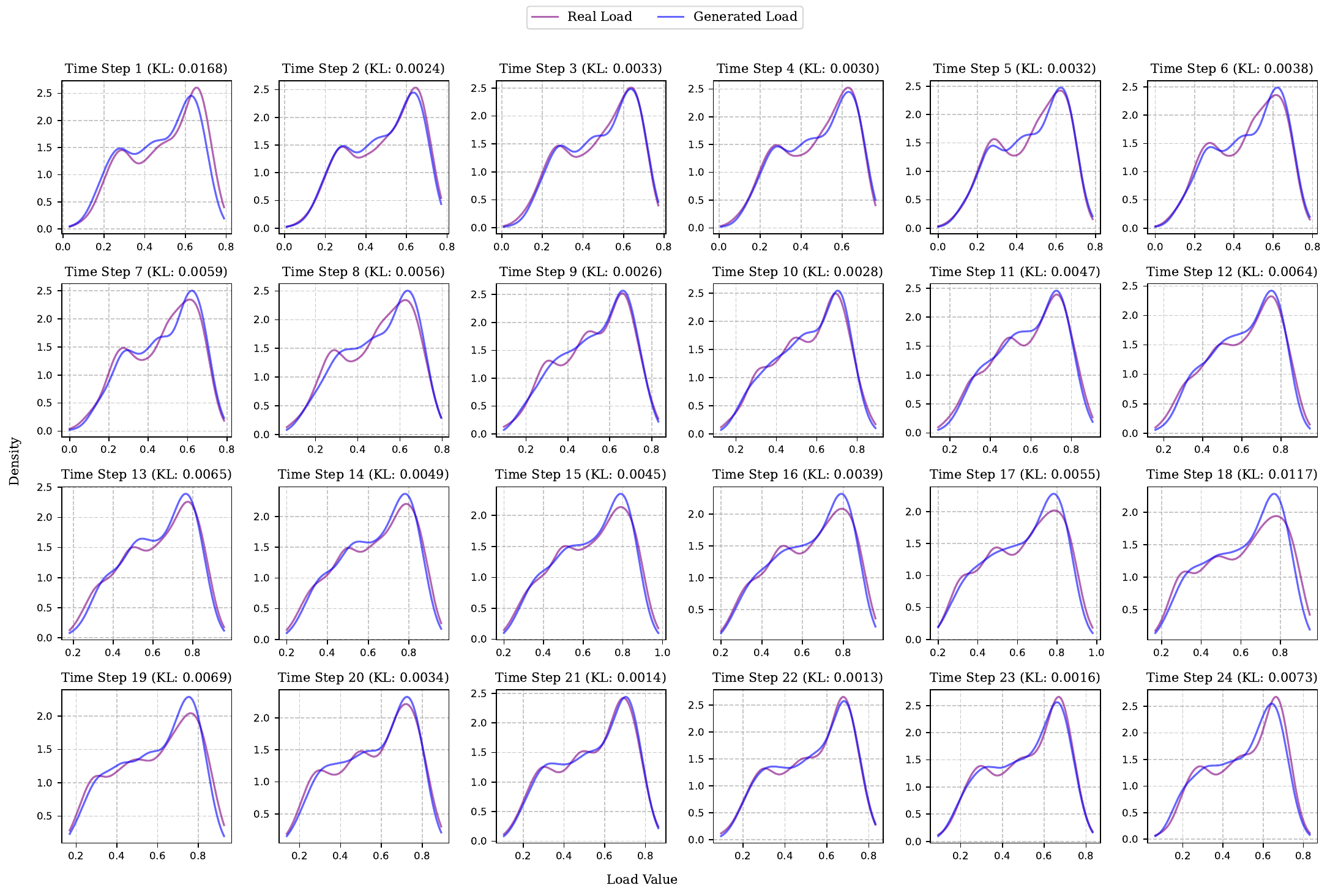} 
    \captionsetup{justification=raggedright, singlelinecheck=false, font=small} 
    \caption{KDE and KL divergence results across 24 time steps for the total test samples.}
    \label{fig:wholepage}
\end{figure*}

\subsection{Evaluation of Generated Load Curves}
Fig. \ref{figure: load curves} shows load curves generated by the diffusion model, obtained by averaging the generated curves. The MSE values are 0.0007, 0.0007, 0.0012, and 0.0004, respectively. Additionally, among the $4\times 24$ time points in Fig. \ref{figure: load curves}, the maximum deviation between the actual and generated curves is 0.0772, which is less than half of the AW value under the 95\% PINC in Table \ref{table: Evaluation Criteria}. It can be inferred that within the 95\% PINC interval, this maximum deviation point also falls within the generated interval. Fig. \ref{figure: load curves} demonstrates that the generated curve accurately captures the overall trend of the actual load, which showcases the diffusion model's significant level of accuracy. 

Moreover, we used KDE to construct the probability distributions of the generated and actual curves at 24-time points. We calculated the KL divergence between the distributions, illustrated in Fig \ref{fig:wholepage}. Among the 24 time steps, the maximum KL divergence is 0.0168 at the time step $1$. According to \cite{KLdivergence}, when the KL divergence is less than 0.05, the two distributions can be considered very similar. Therefore, through Fig. \ref{figure: load curves} and \ref{fig:wholepage}, we can see that the diffusion model not only generates curves but also captures the distribution of the generated curves. Compared to general prediction models such as LSTM and Transformer, the diffusion model demonstrates probabilistic capabilities that typical prediction models do not possess.

\begin{table}[!t]
\centering
\renewcommand{\arraystretch}{1.3}
\setlength{\tabcolsep}{7pt}
\caption{The Evaluation Criteria Results For Different Attention Mechanisms}
\label{table: attention results}

\begin{tabular}{@{}c|l|c|c|c}
\toprule
\multicolumn{1}{c}{\textbf{PINC}} & \multicolumn{1}{c}{\textbf{Attention Type}} & \multicolumn{1}{c}{\textbf{ACE}} & \multicolumn{1}{c}{\textbf{AW}} & \multicolumn{1}{c}{\textbf{\(\overline{\mathbf {S}}^{(\boldsymbol{\alpha})}\)}} \\
\midrule
\multirow{3}{*}{80\%}    & Global                        & 3.32\%          & 0.1559          & -0.0711 \\
                         & Multi-scale                   & 3.18\%          & 0.1673          & -0.0736 \\
                         & \textbf{TMSAB} & \textbf{2.04\%} & \textbf{0.1534} & \textbf{-0.0682} \\
\midrule            
\multirow{3}{*}{90\%}    & Global                        & 2.15\%          & 0.2008          & -0.0481 \\
                         & Multi-scale                   & 2.33\%          & \textbf{0.1939} & -0.0465 \\
                         & \textbf{TMSAB} & \textbf{2.01\%} & 0.1941          & \textbf{-0.0449} \\
\midrule
\multirow{3}{*}{95\%}    & Global                        & \textbf{1.15\%} & 0.2196          & -0.0282 \\
                         & Multi-scale                   & 1.72\%          & 0.2229          & -0.0293 \\
                         & \textbf{TMSAB} & 1.22\%          & \textbf{0.2172} & \textbf{-0.0277} \\
\bottomrule
\end{tabular}
\end{table}

\subsection{Comparison of Three Different Attention Mechanisms}
Table \ref{table: attention results} presents the numerical results obtained from three different attention mechanisms. When PINC is set to 80\%, TMSAB achieved the best results across all three metrics, leading the second-best results by 1.14\%, 0.0139, and 0.0029, respectively. With PINC at PINC = 90\%, it only slightly lags behind Multi-scale attention in AW by 0.0002 while outperforming the other two attention mechanisms in the remaining metrics. At 95\% PINC, the ACE metric is slightly lower than that of global attention by 0.07\%.

\begin{figure}[!t]
  \centering
  \includegraphics[width=5.49in]{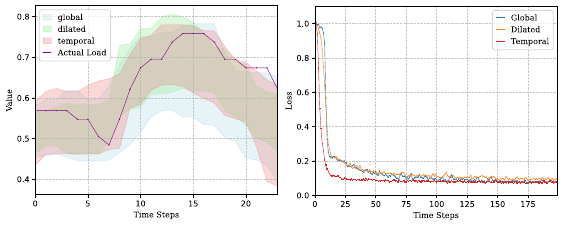}
  \captionsetup{justification=raggedright, singlelinecheck=false, font=small} 
  \caption{Left: Probabilistic forecasting results of three different attention mechanisms under 95\% PINC; Right: Training loss of different attention mechanisms.}
  \label{figure Attention loss}
\end{figure}

Additionally, Fig. \ref{figure Attention loss} displays the results of random samples at 95\% PINC and the training loss for the three attention mechanisms. It is evident that TMSAB achieves superior results in both ACE and AW and demonstrates a faster convergence rate compared to the other two attention mechanisms.

\section{Conclusion}
In this paper, we introduce \modelname, a novel denoising diffusion model designed to achieve PDALF. To enhance \modelname, we developed an attention block, TMSAB, based on multi-scale attention, which integrates both positional and temporal information within the sequence. Experimental data demonstrate that \modelname\ effectively approximates the complex distribution of real load time-series data, with the inclusion of TMSAB significantly improving prediction accuracy. Moreover, this generative approach, as opposed to direct prediction, not only adapts well to forecasting tasks but also provides a fresh perspective for other tasks in power systems.

\bibliographystyle{elsarticle-num} 
\bibliography{elsa.bib}

\end{document}